\documentclass{article}
\usepackage{amsmath}

\input{tcilatex}
\begin{document}

\title{\textbf{2D Schr\"{o}dinger Equation with Mixed Potential in
Noncommutaive Complex space }}
\date{}
\author{Slimane Zaim and Hakim Guelmamene  \\
D\'{e}partement de Physique ,\\
Facult\'{e} des Sciences de la Mati\'{e}re, Universit\'{e} Batna1, Algeria. 
\\
}
\maketitle

\begin{abstract}
We study the 2D Schrdinger equation for Hydrogen atom with the lenear plus
Harmonic Potentials in noncommutative complex space, using the Power-series
expansion method and perturbation theory. Hence we can say that the Schr\"{o}%
dinger equation in noncommutative complex space describes to the particles
with spin (1/2) in an external uniform magnitic field. Where the
noncommutativity play the role of magnetic field with created the total
magnetic moment of particle with spin 1/2, who in turn shifted the spectrum
of energy. Such effects are similar to the Zeeman splitting in a commutative
space. It is shown that on the noncommutative space the degeneracy of the
levels is lifted completely.


Pacs numbers:11.10.Nx, 32.30-r, 03.65-w
\end{abstract}

\section{introduction:}

The non-commutative quantum mechanics is motivated by the natural extension
of the commutation relations between position and momentum, by imposing
further commutation relations between position coordinates themselves. the
non-commutativity of position coordinates immediately implies a set of
uncertainty relations between position coordinates analogous to the
Heisenberg uncertainty relations between position and momentum; namely:

\begin{equation}
\Delta \hat{x}_{i}\Delta \hat{x}_{j}\succeq \frac{\theta _{ij}}{2}  \tag{1}
\end{equation}

Using the new noncommutative coordinate operators $\hat{x}$ satisfying the
commutation relations:

\begin{equation}
\left[ \hat{x}_{i},\hat{x}_{j}\right] ==i\theta _{ij}  \tag{2}
\end{equation}

Where $\hat{x}^{i}=x^{i}-\frac{\theta ^{ij}}{2}p_{j}$ and the parameter $%
\theta ^{ij}$ is an antisymmetric real matrix of dimension square length in
the noncommutative canonical-type space. Where the ordinare product betwenn
functions in noncommutive space is remplaced by the\ star product, where
the\ star product between two arbitrary functions $f\left( x\right) $ and $%
g\left( x\right) $ is given by:

\begin{equation}
f\left( x\right) \star g\left( x\right) =f\left( x\right) .g\left( x\right)
+\left. \exp \left( \frac{i}{2}\theta ^{ij}\partial _{i}\acute{\partial}%
_{j}\right) f\left( x\right) g\left( \acute{x}\right) \right\vert _{x=\acute{%
x}}  \tag{3}
\end{equation}

We observed that the result of eq. $(1)$ satisfied by applying the notion of
the star product represented by eq.$(2)$. The important problems of
noncommutative quantum mechanics is to solve the Schrodiger equation with
physical potentials in noncommutative space. Actually there are many
attempts to study noncommutative space but in two dimensional complex space
the study is a limited attempts $\left[ 1,2\right] $. The study of the Schr%
\"{o}dinger equation for Hydrogen atom with a linear plus Harmonic
potentials has proved to be fruitful and many papers have been published $%
[3-12].$ In this work we study an important contribution to the
non-commutative geometry approach to the hydrogen atom with \ linear plus
Harmonic potentials. Our goal is to solve the Schr\"{o}dinger equation for
Hydrogen atom with linear plus Harmonic potentials in a non-commutative
complex space up to first-order of the non-commutativity parameter using the
Moyal product , we show that the non-commutative value of the energy levels
indicates the splitting of $1s$ states.

This paper is organized as follows. In section $2$ we derive the deformed $%
2D $ Schr\"{o}dinger equation for Hydrogen atom with a linear plus Harmonic
potentials in noncommutative space, we solve this equation and obtain the
non-commutative modification of the energy levels. In section $3$ we derive
the deformed $2D$ Schr\"{o}dinger equation for Hydrogen atom with a linear
plus Harmonic potentials in noncommutative complex space, we solve this
equation and obtain the non-commutative modification of the energy levels.
Finally, in section $4$, we draw our conclusions.

\section{Noncommutative Schr\"{o}dingre equartion with mixed potential}

The $2D$ Schr\"{o}dinger equation with the central potential $V\left( \hat{r}%
\right) $, in noncommutative space, which have the following form:

\begin{equation}
\left( -\frac{1}{2m}\Delta +V\left( r\right) \right) \ast \Psi \left(
r,\varphi \right) =\hat{E}\Psi \left( r,\varphi \right)  \tag{4}
\end{equation}

Expanding the $\ast $ product for equ.(3) we get

\begin{equation*}
\left( -\frac{1}{2m}\Delta +V\left( \hat{r}\right) \right) \Psi \left(
r,\varphi \right) =\hat{E}\Psi \left( r,\varphi \right)
\end{equation*}

where the $V\left( \hat{r}\right) $ ( coulomb plus linear plus oscilator
harmonic) in nonocommutative space as :

\begin{equation}
V\left( \hat{r}\right) =\frac{c}{\hat{r}}+a\hat{r}+b\hat{r}^{2}  \tag{5}
\end{equation}

and the deformed of $\hat{r}$ to first order in the nonocommutative
parameter $\theta $ as:

\begin{equation}
\hat{r}=r-\frac{\theta \cdot L}{2r}+\mathcal{O}\left( \Theta ^{2}\right) 
\tag{6}
\end{equation}

Then the deformed of the central potential $V\left( \hat{r}\right) $ in eq.$%
(5)$ up to $\mathcal{O}\left( \Theta ^{2}\right) $ as:

\begin{eqnarray}
V\left( \hat{r}\right)  &=&ar+br^{2}+\frac{c}{r}-\left( \frac{a}{2}+b\right)
\theta \cdot L+\frac{c\theta \cdot L}{2r^{3}}+\mathcal{O}\left( \Theta
^{2}\right)   \notag \\
&=&V\left( r\right) +V_{\text{NC}}^{\theta }\left( r\right) +\mathcal{O}%
\left( \Theta ^{2}\right)   \TCItag{7}
\end{eqnarray}

where

\begin{equation}
V_{\text{NC}}^{\theta }=-\left( \frac{a}{2}+b\right) \theta \cdot L+\frac{%
c\theta \cdot L}{2r^{3}}  \tag{8}
\end{equation}

wthere the first term ( not depeneds on the distance $r)$, which is similar
to the interaction between the spin and angular momentum, where the
non-commutativity leads the role of the spin and the second term is
represent the dipole-dipole interaction and is the leading-order
perturbation . For simplicity, first of all, we choose the coordinate system 
$\left( x,y,z\right) $ so that $\theta ^{xy}=-\theta ^{yx}=\theta \delta
^{xy}=\theta _{z},$ such that $\theta \cdot L=\theta L_{z}$ and assume that
the other components are all zero ,then the noncommutative Hamiltonian up to 
$\mathcal{O}\left( \Theta ^{2}\right) $ as the form:

\begin{equation}
\hat{H}=\hat{H}_{O}+H_{\text{pert}}^{\theta }  \tag{9}
\end{equation}

where

\begin{equation}
\hat{H}_{O}=H_{O}-\theta \left( \frac{a}{2}+b\right) \theta L_{z}\text{\ and
\ }H_{\text{pert}}^{\theta }=\frac{c\theta L_{z}}{2r^{3}}  \tag{10}
\end{equation}

To investigate the modification of the energy levels by eq. $(10)$, we use
the first-order perturbation theory. The spectrum of $\hat{H}_{0}$ and the
corresponding wave functions are well-known and given by $\left[ 3\right] $:

\begin{equation}
\left\{ 
\begin{array}{c}
\Phi \left( \varphi \right) =\exp \left( \pm im\varphi \right) \text{ where }%
m=0;1,2... \\ 
R_{m}^{n}\left( r\right) =\exp \left( -\frac{ar+br^{2}}{2\sqrt{b}}\right) 
\underset{p=0}{\dsum\limits^{n}}a_{p}r^{p+\delta }%
\end{array}%
\right. ,  \tag{11}
\end{equation}

and

\begin{equation}
\hat{E}_{n,m}=2\sqrt{b}\left( 1+m+n\right) -\frac{\theta }{2}\left(
a+2b\right) m  \tag{12}
\end{equation}

Now, the correction to the energy to first order in $\theta $ is:

\begin{eqnarray}
\Delta E_{n,m}^{\theta } &=&\left\langle n\right\vert H_{\text{pert}%
}^{\theta }\left( r\right) \left\vert n\right\rangle =\theta m\int \Psi
^{\left( n\right) \ast }\left( r\right) \frac{c}{2r^{3}}\Psi ^{\left(
n\right) }\left( r\right) rdrd\varphi  \notag \\
&=&c\pi \theta m\int R^{\left( n\right) \ast }\left( r\right) \frac{1}{r^{2}}%
R^{\left( n\right) }\left( r\right) dr  \TCItag{13}
\end{eqnarray}

where:

\begin{eqnarray}
\int R^{\left( n\right) \ast }\left( r\right) \frac{1}{r^{2}}R^{\left(
n\right) }\left( r\right) dr &=&\underset{p=0}{\dsum\limits^{n}}C_{p}\int
r^{p+2\delta -2}\exp \left( -\frac{ar+br^{2}}{2\sqrt{b}}\right) dr  \notag \\
&=&\exp \left( \frac{a^{2}+2ba}{4b\sqrt{b}}\right) \underset{p=0}{%
\dsum\limits^{n}}C_{p}b^{-\frac{n+2\delta }{2}}\Gamma \left( n+2\delta ,-%
\frac{a}{2\sqrt{b}}\right)   \TCItag{14}
\end{eqnarray}

where

\begin{equation}
C_{p}=\underset{k=0}{\dsum\limits^{p}}a_{k}a_{p-k}  \tag{15}
\end{equation}

Then the modification of the energy levels is given by:

\begin{equation}
\Delta E_{n,m}^{\theta }=\theta \left( c\pi \exp \left( \frac{a^{2}+2ba}{4b%
\sqrt{b}}\right) \underset{p=0}{\dsum\limits^{n}}C_{p}b^{-\frac{n+2\delta }{2%
}}\Gamma \left( n+2\delta ,-\frac{a}{2\sqrt{b}}\right) \right) m  \tag{16}
\end{equation}

We have show that the non-commutative energy correction is proportional to
the quantum number $m$ , such are similar to the Zeeman effects and they
remove the degeneracy with respect to $m.$

\section{ Schr\"{o}dingre equartion with mixed potential in\ noncommutative
complex space}

In the complex coordinate $\left( z,\bar{z}\right) $ , the$\ $ radial wave
function $\Psi \left( r,\varphi \right) $ will be written to the form:

\begin{equation}
\Psi \left( r,\varphi \right) =\Psi \left( z,\bar{z}\right) =\frac{%
R_{m}\left( z,\bar{z}\right) }{\left( z\bar{z}\right) ^{1/4}}\Phi \left(
\varphi \right) ,\text{ \ \ \ \ }z\bar{z}=r^{2}  \tag{17}
\end{equation}

where the energy levels and the corresponding radial wave functions are
well-known and given by $\left[ 3,8\right] $:

\begin{equation}
.R_{m}\left( z,\bar{z}\right) =\exp \left( p_{m}\left( z,\bar{z}\right)
\right) \underset{p=0}{\dsum\limits^{n}}a_{n}\left( z\bar{z}\right) ^{\frac{%
p+\delta }{2}}  \tag{18}
\end{equation}

and

\begin{equation}
E_{n}=2\sqrt{b}\left( 1+m+n\right)   \tag{19}
\end{equation}

where $p_{m}\left( z,\bar{z}\right) $ is given by$:$

\begin{equation}
p_{m}\left( z,\bar{z}\right) =\alpha \sqrt{z\bar{z}}+\frac{1}{2}\beta z\bar{z%
}  \tag{20}
\end{equation}

and $\alpha $ , $\beta $ and $\delta $ are given by $:$

\begin{equation}
\left\{ 
\begin{array}{cc}
\beta ^{2}=b, & 2\alpha \beta =a,%
\end{array}%
\right.   \tag{21}
\end{equation}

if we choose $\beta =-\sqrt{b\text{ }}$ and $\delta =\frac{1}{2}+m$ $.$The
exact solution for $n=0$ and $n=1$ are given by $\left[ 3,8\right] $:

\begin{equation}
\left\{ 
\begin{array}{ccc}
R_{m}^{\left( 0\right) }=a_{0}\left( z\bar{z}\right) ^{\delta /2}\exp \left(
-\frac{a\left( z\bar{z}\right) ^{1/2}+bz\bar{z}}{2\sqrt{b}}\right)  & \text{,%
} & E_{0,m}=2\sqrt{b}\left( 1+m\right)  \\ 
R_{m}^{\left( 1\right) }=\left( a_{0}+a_{1}\sqrt{z\bar{z}}\right) \left( z%
\bar{z}\right) ^{\delta /2}\exp \left( -\frac{a\left( z\bar{z}\right)
^{1/2}+bz\bar{z}}{2\sqrt{b}}\right)  & \text{,} & E_{1,m}=2\sqrt{a}\left(
2+m\right) 
\end{array}%
\right.   \tag{22}
\end{equation}

\bigskip In the noncommutative complex space we can show that, in the first
order of the parameter $\theta $: 
\begin{eqnarray}
\hat{z}\widehat{\bar{z}} &=&z\bar{z}-\theta \left( L_{z}-1\right)   \notag \\
\widehat{\bar{z}}\hat{z} &=&z\bar{z}-\theta \left( L_{z}+1\right)  
\TCItag{23} \\
\hat{p}^{2} &=&4p_{z}p_{\bar{z}}=p^{2}  \notag
\end{eqnarray}

Then the noncommutative Hamiltonian $\hat{H}$ associated by the mixed
potential as given by$\left[ 2\right] $:

\begin{equation}
\hat{H}=\frac{2}{m}p_{\hat{z}}p_{\widehat{\bar{z}}}+V\left( \hat{z},\widehat{%
\bar{z}}\right)   \tag{24}
\end{equation}

where the deformed potential $V\left( \hat{z},\widehat{\bar{z}}\right) $ in
the noncommutative complex coordinate $\left( \hat{z},\widehat{\bar{z}}%
\right) $ as:

\begin{equation}
\text{ }V\left( \widehat{r}\right) =\left( 
\begin{array}{cc}
V\left( \hat{z},\widehat{\bar{z}}\right) =a\sqrt{\hat{z}\widehat{\bar{z}}}+b%
\hat{z}\widehat{\bar{z}}+\frac{c}{\sqrt{\hat{z}\widehat{\bar{z}}}} & 0 \\ 
0 & V\left( \widehat{\bar{z}},\hat{z}\right) =a\sqrt{\widehat{\bar{z}}\hat{z}%
}+b\widehat{\bar{z}}\hat{z}+\frac{c}{\sqrt{\widehat{\bar{z}}\hat{z}}}%
\end{array}%
\right)   \tag{25}
\end{equation}

Then  $V\left( \hat{z},\widehat{\bar{z}}\right) $ and $V\left( \widehat{\bar{%
z}},\hat{z}\right) $ as follows:

\begin{eqnarray}
V\left( \hat{z},\widehat{\bar{z}}\right)  &=&V\left( r\right)
+V_{NC}^{+}\left( r\right)   \TCItag{26} \\
V\left( \widehat{\bar{z}},\hat{z}\right)  &=&V\left( r\right)
+V_{NC}^{-}\left( r\right)   \TCItag{27}
\end{eqnarray}

where

\begin{eqnarray}
V_{NC}^{\pm }\left( r\right)  &=&-\theta \left( \frac{a}{2}+b-\frac{c}{2r^{3}%
}\right) \left( L_{z}\mp 1\right)   \notag \\
&=&-\theta \left( \frac{a}{2}+b\right) \left( L_{z}-2s_{z}\right) +\theta 
\frac{c\left( L_{z}-2s_{z}\right) }{2r^{3}},s_{z}=\pm \frac{1}{2} 
\TCItag{28}
\end{eqnarray}

The noncommutative Hamiltonian in eq.$(24)$ as:

\begin{equation}
\hat{H}=H_{O}-\theta \left( \frac{a}{2}+b\right) \left( L_{z}-2s_{z}\right)
+\theta \frac{c\left( L_{z}-2s_{z}\right) }{2r^{3}}  \tag{29}
\end{equation}

where the first term $H_{O}$ is \ the ordinary Hamiltonian of hydrogen atom
with linear plus Harmonic potentials , the second term that not depeneds on
the distance $r$, wisch is simelar to the interaction between the spin and
angular momentum and the therd term is represent the dopol-dipol iteraction
and is the leading-order perturbation . Then the noncommutative Hamiltonian
in eq.$(29)$ we can rewiteen as the form:

\begin{equation}
\hat{H}=\hat{H}_{O}+H_{\text{pert}}^{\theta }  \tag{30}
\end{equation}

where

\begin{equation}
\hat{H}_{O}=H_{O}-\theta \left( \frac{a}{2}+b\right) \left(
L_{z}-2s_{z}\right) \text{ \ }  \tag{31}
\end{equation}

and \ 

\begin{equation}
H_{\text{pert}}^{\theta }=\theta \frac{c\left( L_{z}-2s_{z}\right) }{2r^{3}}
\tag{32}
\end{equation}

To investigate the correction of the energy levels by eq. $(32)$, we use the
first-order perturbation theory. The spectrum of $\hat{H}_{0}$ and the
corresponding eigenfunctions can be written as:

\begin{eqnarray}
R_{m}^{+}\left( z,\bar{z}\right) &=&\exp \left( p_{m}\left( z,\bar{z}\right)
\right) \underset{n=0}{\sum }a_{n}\left( z\bar{z}\right) ^{\frac{n}{2}%
+\delta }\oplus \mid +\succ  \TCItag{32} \\
&&\text{and}  \notag \\
R_{m}^{-}\left( z,\bar{z}\right) &=&\exp \left( p_{m}\left( z,\bar{z}\right)
\right) \underset{n=0}{\sum }a_{n}\left( z\bar{z}\right) ^{\frac{n}{2}%
+\delta }\oplus \mid -\succ  \TCItag{33}
\end{eqnarray}

where the energy levels are given by:

\begin{equation}
\hat{E}_{n;m}^{\pm }=2\sqrt{b}\left( 1+m+n\right) -\theta \left( \frac{a}{2}%
+b\right) \left( m\mp 1\right)  \tag{34}
\end{equation}

Now to obtain the correction energy levels as a result of the eq.(31), we
use perturbation theory. To calculate $\Delta E_{n,m}^{\theta \pm }$\
associate withe spin up and spin down , we use the radial function in
equations $(32)$ and $(33)$, to obtain: 
\begin{eqnarray}
\Delta E_{n,m}^{\theta \pm } &=&\left\langle n\right\vert H_{\text{pert}%
}^{\theta }\left( r\right) \left\vert n\right\rangle =\frac{\theta }{2}%
\left( m\mp 1\right) \int \Psi ^{\left( p\right) \ast }\left( z,\bar{z}%
\right) \frac{c}{z\bar{z}\sqrt{z\bar{z}}}\Psi ^{\left( p\right) }\left( z,%
\bar{z}\right) rdr  \notag \\
&=&\theta \pi \left( m\mp 1\right) \int R^{\left( p\right) \ast }\left(
r\right) \frac{c}{r^{3}}R^{\left( p\right) }\left( r\right) rdr  \TCItag{35}
\end{eqnarray}

The non-commutative modification of the energy levels to the stationary
state $\Delta E_{0,m}^{\theta \pm }$ to the first order in $\theta $ as
given by: 
\begin{eqnarray}
\Delta E_{0,m}^{\theta \pm } &=&\pi \left( m\mp 1\right) \theta \overset{%
+\infty }{\underset{0}{\dint }}\left[ a_{0}r^{\delta }\exp \left( -\frac{%
ar+br^{2}}{2\sqrt{b}}\right) \right] ^{2}\frac{c}{r^{3}}rdr  \notag \\
&=&\theta \left( m\mp 1\right) C_{0}  \TCItag{36}
\end{eqnarray}

\bigskip where :

\begin{eqnarray}
C_{0} &=&\pi ca_{0}^{2}\overset{+\infty }{\underset{0}{\dint }}r^{2\delta
-2}\exp \left( -\frac{ar+br^{2}}{\sqrt{b}}\right) dr  \notag \\
&=&\exp \left( \frac{a^{2}+2ba}{4b\sqrt{b}}\right) a_{0}^{2}b^{-\delta
}\Gamma \left( 2\delta ,-\frac{a}{2\sqrt{b}}\right)   \TCItag{37}
\end{eqnarray}

The noncommutative energy levels to the stationary state as:

\begin{equation}
\hat{E}_{0,m}^{\pm }=2\sqrt{b}\left( 1+m\right) +\left( C_{0}-\frac{a}{2}%
-b\right) \theta \left( m\mp 1\right)   \tag{38}
\end{equation}

We show that the non-commutative value of the energy levels indicates the
splitting of $1s$ states in two levels. Also we can say that the Lamb shift
is actually induced by the space non-commutativity which plays the role of a
magnetic field and spin in the same moment (Zemann effect).

Now, the noncommutative energy levels to the excitation state $n,$ to first
order in $\theta $ as:%
\begin{eqnarray}
\hat{E}_{n,m}^{\pm } &=&2\sqrt{b}\left( 1+m+n\right) +\theta \left( a-\frac{b%
}{2}\right) \left( m\pm 1\right) -2\left( m\pm 1\right) \theta \times  
\notag \\
&&\overset{+\infty }{\underset{0}{\dint }}\left[ \left(
a_{0}+a_{1}r^{2}+...a_{p}r^{2p}\right) \right] ^{2}\left[ r^{\delta -1}\exp
\left( -\frac{ar+br^{2}}{2\sqrt{b}}\right) \right] ^{2}dr  \notag \\
&=&2\sqrt{b}\left( 1+m+n\right) +\theta \left( a-\frac{b}{2}-2A_{n}\right)
\left( m\pm 1\right)   \TCItag{39}
\end{eqnarray}

\bigskip where

\begin{equation}
A_{n}=\exp \left( \frac{a^{2}+2ba}{4b\sqrt{b}}\right) \underset{k=0}{%
\dsum\limits^{n}}C_{p}b^{-\frac{k+2\delta }{2}}\Gamma \left( k+2\delta ,-%
\frac{a}{2\sqrt{b}}\right)   \tag{40}
\end{equation}

This result is very important, as a possible means of introducing electron
spin we replace $l=\pm \left( j+1/2\right) $and $n\rightarrow n-j-1-1/2$,
where $j$ is the quantum number associated to the total angular momentum.
Then the $l=0$ state has the same total quantum number $j=1/2$. In this case
the non-commutative value of the energy levels indicates the splitting of $1s
$ states.

\section{Conclusion}

In this paper we started from quantum Hydrogen atom with the linear plus
Harmonic potentials in a canonical non-commutative complex space, we have
derived the defermed Schrodinger equation. Using the power series expansion
method to solving and we found that the noncommutative energy levels, the
energy levels are shifted to $\left( 2j+1\right) $ levels ,it acts here like
a Lamb shift in Dirac theory. This proofs that the non-commutativity has an
effect similar to the Zeeman effects which iduced by the magnetic field $%
\left[ 13\right] $ , where the non-commutativity leads the role of the
magnetic field.

\bigskip

\end{document}